\documentclass[aps,11pt,oneside,onecolumn,groupaddress]{revtex4}

\usepackage{amsmath}
\usepackage{amsfonts}
\usepackage{amssymb}
\usepackage{graphicx}
\usepackage{graphics}
\usepackage{mathcomp}
\usepackage[usenames,dvipsnames]{color}
\usepackage{ulem}      
\usepackage{textcomp} 
\usepackage{mathcomp} 

\newcommand{\one}{{\hat{I}}}
\newcommand{\I}{{\hat{I}}}
\newcommand{\Y}{{\hat{Y}}}
\newcommand{\Z}{{\hat{Z}}}
\newcommand{\X}{{\hat{X}}}

\newcommand{\uno}{\color{black}}    
\newcommand{\due}{\color{black}}   
\newcommand{\tre}{\color{black}}  
\newcommand{\gen}{\color{black}} 

\begin{document}

\title{Quantum Key Distribution using a Two-way Quantum Channel}

\author{Marco Lucamarini}
\email{marco.lucamarini@unicam.it}
\affiliation{School of Science and Technology, University of
Camerino, via Madonna delle Carceri, 9 - 62032 Camerino, Italy}

\author{Stefano Mancini}
\email{stefano.mancini@unicam.it}
\affiliation{School of Science and Technology, University of
Camerino, via Madonna delle Carceri, 9 - 62032 Camerino, Italy}

\begin{abstract}
{We review a quantum key distribution protocol, recently proposed
by us, that makes use of a two-way quantum channel. We provide a
characterization of such a protocol from a practical perspective,
and consider the most relevant {\tre individual-particle}
eavesdropping strategies against it. This allows us to compare its
potentialities with those of {\tre the} standard {\tre BB84}
protocol which uses a one-way quantum channel.}
\end{abstract}


\maketitle

\newpage

\tableofcontents

\newpage

\section{Introduction}

Since the seminal works by Bennett and Brassard \cite{BB84} and by
Ekert \cite{Ek91} Quantum Key Distribution (QKD) has made
impressive progresses~\cite{Gisin02}, which can be roughly grouped
into two main categories: on  the one side there are theoretical
progresses, among which the unconditional security of QKD is the
most relevant one \cite{May96,LC99,SP00,Ben-Or05,Ren07}; on the
other side there are experimental progresses, almost exclusively
in the field of quantum optics, which recently led to long-haul
and high-rate QKD experiments~\cite{Stu02,Sch07,Urs07,Tak07}. The
relevance of these advances promoted QKD as the most applicative
research of quantum information, and also triggered the start-up
of companies based on this technology~\cite{Cit1}.

Recent research on QKD is mainly focused on closing the existing
gap between the perfect theory of the unconditional security
proofs and the imperfect application of such a theory in the real
world. This originated the definition of \textit{practical
QKD}~\cite{Lut99,Lut00} which deals with the proper modeling of
devices like photon sources, light modulators and single-photon
detectors~\cite{Dus06,Lo07}. By consequence, any new proposal
relevant to the field of QKD can not set aside the practicality
issue. This is even truer if one looks at the quantum hacking
strategies based on practical imperfections recently accomplished
by the group of Trondheim~\cite{Mak06}.

In this paper we want to characterize from a practical point of
view a novel form of QKD based on a two-way quantum channel.
Two-way channels are already used in QKD either in those setups
based on a {Plug-and-Play} configuration~\cite{Muller1996} or in
those using the {Cascade} error correction
procedure~\cite{Brassard1994}. However these two examples do not
represent a two-way \textit{quantum} channel: in the first case
only the backward channel ought to be considered quantum; in the
second case neither the forward nor the backward channel are
quantum.

The first input to \textit{two-way QKD} came from Bostr\"{o}m and
Felbinger's Ping-Pong protocol~\cite{BF02}, where the question of
the security of Quantum Dense Coding~\cite{BW92} was posed in
terms of ``deterministic and secure direct communication in
presence of entanglement''. {\gen The Ping-Pong protocol was
demonstrated insecure~\cite{Cai03,Woj03} but a revised, more
secure, version was soon after proposed in~\cite{CL04} (see
also~\cite{DRC+04,Woj05,DRC+05}). }

Moving from these initial steps novel two-way schemes
\textit{without entanglement} were presented {\gen
in~\cite{CL04b}, in~\cite{DL04}} and then by us in~\cite{Luc05}.
This last scheme, named ``LM05'', had the merit of being more
feasible and secure than the protocol in~\cite{BF02} {\gen and it
came with the first security proof against non-trivial
individual-particle attacks}. {\gen As a matter of} fact its
experimental realization {\gen was} proposed~\cite{Luc05a}, proved
in principle~\cite{Cer06} and recently realized in the third
telecom window~\cite{Kum08}. Also, the protocol {\gen was}
generalized to a higher number of states~\cite{Shaari06,SSK+09}
and to higher-dimensional Hilbert
spaces~\cite{Shaari08,Eusebi08,Piran08}.

{\tre Despite these improvements, a complete realization of the
LM05 still remains an experimental challenge. The main problem is
that in the original protocol both the forward and the backward
channels are tested by the users to guarantee the security of the
protocol. In particular the test of the backward channel requires
that with a certain probability one of the users (Alice) prepares
a qubit in exactly the same physical state as the one prepared by
the other user (Bob)~\cite{Luc05}. This is difficult in practice
because all the relevant degrees of freedom of the qubit
(wavelength, timing, bandwidth, etc.) must be accurately
controlled by Alice.}

{\tre In this work we modify the original LM05 protocol by
removing the test of the backward channel and show that the
resulting protocol is still secure against several single-particle
attacks by the eavesdropper (Eve).}
We also provide the state-of-the-art of the so called
``deterministic QKD''~\cite{footnote1} with a two-way quantum
channel, together with a set of tools which can be used and
further improved towards a final security proof {\tre of the
scheme.} 
%
Throughout the work we compare our \textit{deterministic} protocol
with the standard \textit{non-deterministic}
BB84~\cite{NoteEffBB84}.


\section{The protocol}
\label{sec:2}

\begin{figure}[ht]
\includegraphics[width=12.0cm]{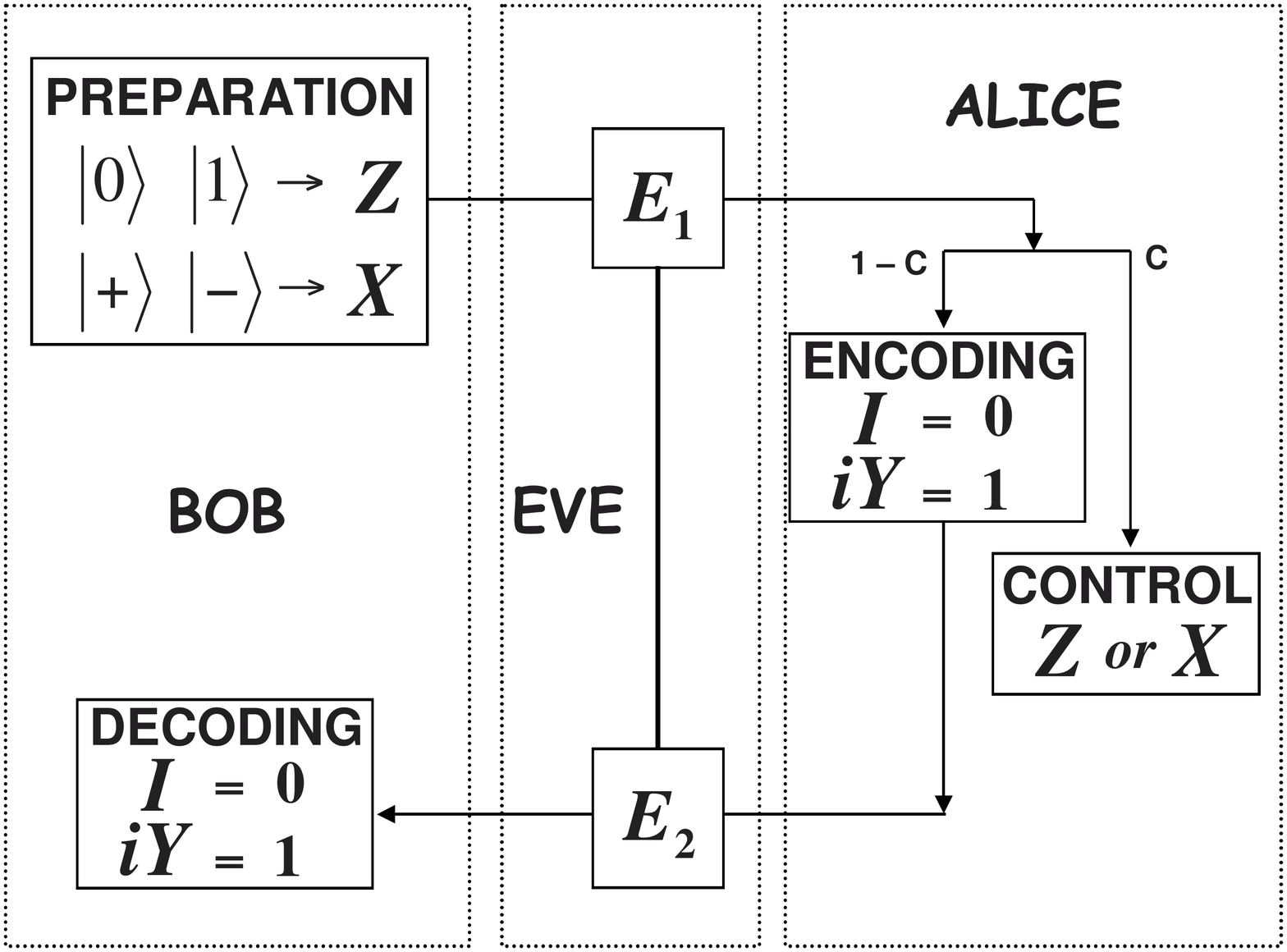}\caption{Schematics
of the practical LM05 protocol. Bob randomly prepares qubits of
$\X$ and $\Z$ bases. With probability (1-c) Alice encodes data on
the qubits (Encoding Mode, EM) and with probability (c) she
measures the qubits (Control Mode, CM). Bob measures the returning
qubits in the same bases they were prepared and finally decodes
Alice's information. The noise check on the backward path, present
in the original protocol~\cite{Luc05}, is removed in the present
version.} \label{fig:1-protocol}
\end{figure}

The practical rendering of the LM05 protocol (see
Fig.\ref{fig:1-protocol}) can be obtained
from the original protocol~\cite{Luc05} by removing the test
performed by the users on the backward channel. Bob prepares a
qubit in one of the four states $|0\rangle$, $| 1\rangle$ (the
Pauli $\Z$ eigenstates), $| +\rangle$, $| -\rangle$ (the Pauli
$\X$ eigenstates), and sends it to his counterpart Alice. With
probability $c$ Alice switches to Control Mode (CM) and uses the
qubit to test the channel noise or, with probability $1-c$, she
switches to Encoding Mode (EM) and uses the qubit to encode a bit
of information. The CM consists in a projective measurement of the
qubit along a basis randomly chosen between $\Z$ and $\X$, in a
way equal to the protocol BB84~\cite{BB84}. In the original
protocol~\cite{Luc05} this step was followed by the preparation of
a new qubit in the same state as the outcome of the measurement;
however this step is removed from the present protocol as it is
not practical to implement. {\tre In fact, to conceal the chosen
modality to Eve, Alice should prepare a qubit in exactly the same
physical state as the one prepared by Bob, i.e. same wavelength,
bandwidth, time-width and intensity. This is not possible without
a sophisticated control of the experimental devices. The lack of a
direct test of the backward channel thus} represents the
fundamental difference respect to the protocol {\tre described
in~\cite{Luc05}}. The EM is a modification of the qubit state
according to one of the following transformations: the identity
operation $\I$, which leaves the qubit unchanged and encodes the
logical `0', or $i\Y \equiv \Z \X$ operation, which flips any of
the qubits prepared by Bob and encodes the logical `1'. After the
encoding step Alice sends the qubit back to Bob who measures it in
the same basis he prepared it; in case of an EM run this feature
allows Bob to \textit{deterministically} infer Alice's operation,
without any need of a basis reconciliation procedure, typical of a
\textit{non-deterministic} setup like BB84~\cite{footnote2}.
Notice that Bob's measurement does not depend on the modality
chosen by Alice (EM or CM): {\tre Bob will perform in any case his
measurement, with a nonzero probability to detect a vacuum pulse
due either to the natural losses of the channel or to a CM run by
Alice or to a specific attack by Eve. After the quantum
communication, Alice will tell on the classical channel which runs
were CM and which were EM.} We also point out that the {\gen
direct test of at least one of the two channels} is a necessary
procedure for any two-way quantum protocol. In fact there exist
attacks, like Trojan-horse~\cite{gisin06}, Quantum
Man-in-the-Middle~\cite{LucPhd05} and
Double-CNOT~\cite{LucPhd05,Shaari06}, described in next
Section~\ref{subsec:DCNOT}, which can not be detected by Alice and
Bob using only the EM runs.

Even if the bases are not revealed for the EM runs, they are
revealed for comparing the data acquired during the CM runs. This,
in complete analogy with the BB84, will provide the users with an
estimate of the noise present on the forward channel in terms of
the QBER (Quantum Bit Error Rate) \textbf{$q_1$}, defined as the
ratio of the number of wrong bits over the number of total bits
coming from CM runs. Beside $q_1$, the users can also estimate a
second QBER \textbf{$Q_{AB}$} which comes from a fraction of the
EM runs, in which Alice reveals on the public channel her
encoding.

While $q_1$ is used to give an upper bound on Eve's information,
$Q_{AB}$ is used to provide an estimate of Alice and Bob mutual
information, $I_{AB}$, with $I_{AB} = H(A)-H(A|B)$ and $H$ the
usual Shannon entropy~\cite{NC00}. In case $A$ is a random binary
variable we have:
\begin{equation}\label{IAB}
    I_{AB}(Q_{AB}) = 1 - H(Q_{AB}).
\end{equation}
Note that there is no definite relation between the two QBERs
$q_1$ and $Q_{AB}$. A precise relation can only be found by
assuming a particular strategy by Eve, as done in~\cite{Luc05} or
by adopting a particular noise model, as we shall do in the
following.
%

Eve's information is given in terms of mutual information between
Eve and Alice, $I_{AE}$, or Eve and Bob, $I_{BE}$. Then the
Csisz\'ar-K\"orner (CK) theorem~\cite{CK78} is used to decide
whether the QKD session is secure or not: Alice and Bob can
establish a secret key (using Forward Error Correction and Privacy
Amplification) if and only if $I_{AB} \geq I_{AE}$ or $I_{AB} \geq
I_{BE}$. Depending on the use of the $I_{AE}$ or $I_{BE}$ in the
CK theorem, it is possible to define respectively the
\textit{secrecy capacity} of the Direct Reconciliation (DR),
$C_s^{DR}$, and of the Reverse Reconciliation (RR), $C_s^{RR}$, as
follows:
{\gen
\begin{eqnarray}\label{CKtheorem}
  C_s^{DR}(Q_{AB},q_1) &=&I_{AB}(Q_{AB})-\overline{I}_{AE}(q_1),\\
  C_s^{RR}(Q_{AB},q_1) &=&I_{AB}(Q_{AB})-\overline{I}_{BE}(q_1),
\end{eqnarray}
where $\overline{I}_{AE}$, $\overline{I}_{BE}$ are the upper
bounds to Eve's mutual information with Alice and Bob
respectively, which as said are functions of $q_1$ only.} Notice
that according to the CK theorem is sufficient that only one of
the secrecy capacities is greater than zero to have a secure
communication.
In order to accomplish a DR, the users must execute the Forward
Error Correction in the direction that goes from Alice to Bob; in
other terms it will be Alice to transmit to Bob the parity bits to
correct the errors, and it will be Bob to correct the errors in
his string to match Alice's string. For the RR the roles of the
users are the opposite.
We remark that to have a tentatively deterministic Direct
Communication~\cite{BF02}, it must be Alice to lead the
reconciliation procedure, because only her knows the random choice
between CM and EM. This means that the security of a deterministic
communication closely depends on the mutual information $I_{AE}$
and on the DR procedure.

The LM05 protocol has already been tested in free-space at the
wavelength of 800nm using the photon polarization as a quantum
carrier of the information~\cite{Cer06,Luc06}. However this choice
is not ideal in optical fibers because birefringence makes the
polarization change randomly. The optimal implementation is to use
an encoding based on the relative phase between two pulses
separated in time. In this way the channel noise seen by the two
pulses is the same, and their relative phase remains stable. The
only left noise is the one coming from a misalignment between
Alice's and Bob's apparatuses. To eliminate this source of noise
one can adopt the mechanism of passive compensation, invented
in~\cite{Martinelli1989} and employed for QKD in the Plug-and-Play
system~\cite{Zbinden1997,Ribordy1998,Ribordy2000,Bethune2000}.
This technique fits very natural with the EM of LM05, thus letting
the implementation reported in~\cite{Kum08}.

In the next sections we consider the most relevant individual
attacks against LM05. Following the approach given
in~\cite{Bennett92} we group the attacks into two main categories:
\textit{zero-loss} and \textit{zero-QBER} attacks. The former,
{\gen analyzed in Section~\ref{sec:0loss}}, do not introduce any
loss in the channel but introduce noise; the latter, {\gen treated
in Section~\ref{sec:0QBER},} introduce losses, but no noise. This
description allows to a certain extent a fair comparison between
one-way and two-way {\gen deterministic} QKD.

\section{Zero-loss eavesdropping}
\label{sec:0loss}


\subsection{Intercept and resend attack - IR}
\label{subsec:IR}

The simplest attack in which Eve acquires information from the
communication channel, thus unavoidably introducing noise in the
same channel, is the \textit{Intercept and Resend} (IR), {\gen
which follows for the} LM05 {\gen protocol} the same steps as for
the BB84 protocol~\cite{BB84,Bennett92}.

In the IR against LM05 Eve measures the photon coming out from Bob
along one basis chosen at random between the two bases used by
Bob, i.e. $\hat{Z}$ and $\hat{X}$~\cite{footnote3}. Then she takes
note of the outcome and forwards to Alice the qubit projected by
her measurement in a certain state $\left| \psi \right\rangle$.
Alice, in EM, will flip or not-flip this state, and sends it back
to Bob. Eve measures on the backward path the state emerging from
Alice box, using the same basis previously used in the forward
path, thus learning the information; finally she forwards the
qubit to Bob.

This attack can be detected through the QBER $q_1$ while running
the CM. It is easy to see that when Eve perchance measures in the
same basis of Alice and Bob, she creates no error in the forward
channel; on the contrary, for different basis, she creates an
error with probability $1/2$. In both cases she acquires full
information about Alice's encoding, causing an average value of
$q_1$ equal to $1/4$.

While IR provides Eve with full knowledge about Alice's encoding it
does not provide her with full knowledge about the result of Bob's
measurement. In fact the first measurement by Eve destroys the
initial state prepared by Bob; furthermore the basis used by Bob
is never revealed in LM05. This entails that Eve can foresees
Bob's result only three times over four, that is when she
perchance measures along the same basis as Bob and when she
measures along the wrong basis but accidentally guesses the
correct final result. This argument, together with the previous
one, implies the following relations for the IR:
\begin{eqnarray}
\nonumber  q_1 &=& 0.25 ,\\
\nonumber  Q_{AE} &=& 0 \Rightarrow I_{AE} = 1 , \\
           Q_{BE} &=& 0.25 \Rightarrow I_{BE} = 1-H(Q_{BE}) \simeq 0.19.
           \label{eq:IR-LM05}
\end{eqnarray}
In the above equations $Q_{AE}$ and $Q_{BE}$ are defined in a way
similar to $Q_{AB}$. For example $Q_{AE}$ can be thought as the
QBER that Alice and Eve would find if they compare their classical
strings which are composed, respectively, by all the encodings
transmitted to Bob and by all the data acquired by Eve by means of
the IR. Recall that the above equations are connected with the
direct (DR) and reverse (RR) reconciliation of the LM05 and
provide the direct and reverse secrecy capacity of the protocol.
The situation described by Eq.\eqref{eq:IR-LM05} can be compared
with that of BB84, where for $q_1=0.25$ the information acquired
by Eve is $I_{AE}=I_{BE}=1/2$~\cite{Gisin02}.

Of course Eve can decide to effect or not to effect IR in a
certain run. In other terms she can effect IR only on a fraction
$\xi\in[0,1]$ of the runs. This modifies Eqs.\eqref{eq:IR-LM05}
into the new ones:
\begin{eqnarray}\label{q1_IR}
\nonumber  q_1 &=& 0.25~\xi ,\\
\nonumber  I_{AE} &=& \xi , \\
           I_{BE} &\simeq& 0.19~\xi.
\end{eqnarray}

As said above, the mutual information between Alice and Bob is a
simple function of the QBER $Q_{AB}$, which is measured by the
users by sacrificing a part of the data collected during the EM.
It is straightforward to realize that for IR the two QBERs of the
LM05 are equal:
\begin{equation}\label{q1qab}
    Q_{AB} = q_1.
\end{equation}
{\gen
Eq.\eqref{q1qab} defines a first relation between the two QBERs
measured in the LM05 protocol. Although this relation is not
general, it deserves attention for it describes quite accurately
the noise model pertaining to an experimental fiber-based setup
like the one in Ref.~\cite{Kum08}, where the phase-drift noise
arising in the optical fiber can be compensated both at the end of
the forward channel and at the end of the backward channel through
the passive Plug-and-Play auto-compensation
technique~\cite{Zbinden1997,Ribordy1998,Ribordy2000,Bethune2000}.
Hence, we will use again the noise model of Eq.\eqref{q1qab} to
settle a homogeneous comparison between the various attacks put
forward by Eve against the LM05 protocol. Of course, other models
of noise are possible (see e.g. Ref.~\cite{Cer06}).

In Fig.~\ref{fig:IRinfos} we summarize what obtained thus far with
the curves pertaining to the mutual information between Alice, Bob
and Eve in LM05.
\begin{figure}[h!]
  \includegraphics[width=12.0cm]{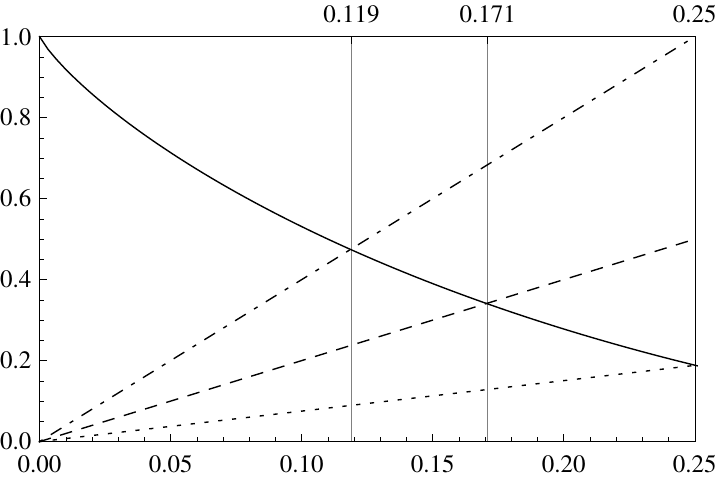}\\
  \caption{Mutual information versus $q_1$ in LM05 and BB84 in case of
  IR attack by Eve.
  $I_{AB}$: solid line;
  $I_{AE}^{LM05}$: dot-dashed line;
  $I_{BE}^{LM05}$: dotted line;
  $I_{AE}^{BB84}=I_{BE}^{BB84}$: dashed line.
  The security thresholds are reported at the top of the frame.}
  \label{fig:IRinfos}
\end{figure}
{\gen In the same figure we also plot the curve pertaining to the
analogous IR attack against BB84 (Section VI.D of~\cite{Gisin02}),
for which $I_{AE}=I_{BE}$.}
It can be seen that the curve $I_{AE}$ in LM05 is always above the
curve of BB84, while the curve $I_{BE}$ is always below it. This,
{\gen according to the CK theorem, Eq.\eqref{CKtheorem},} implies
that the reverse (direct) secrecy capacity of LM05 is always
higher (lower) than that of BB84 for the IR attack. {\gen The
security thresholds obtained from the CK theorem against the IR
attack for LM05 in DR and RR and for BB84, in the noise model of
Eq.\eqref{q1qab}, are given by the intersection points of the
plotted curves, and are explicitly written below:
\begin{eqnarray}
\nonumber  LM05^{DR} &:& \textrm{secure against IR if } q_1<11.9\% \\
\nonumber  LM05^{RR} &:& \textrm{secure against IR if } q_1<25.0\%\\
\nonumber  BB84 &:& \textrm{secure against IR if } q_1<17.1\%
\end{eqnarray}
Note that the BB84 security threshold of $17.1\%$ is higher than
the one pertaining to the optimal individual
eavesdropping~\cite{FGN+97} which amounts to $14.6\%$. In fact the
simple-minded IR attack just described is not optimal either for
BB84 or for LM05.
}
{\uno However they have been used in~\cite{Bennett92} and later
in~\cite{Kum08} to provide direct measurable quantities easily
applicable in the experimental situation.}

\subsection{Non-orthogonal attack - NORT}
\label{subsec:NonOrt}

The IR analyzed above can give Eve an information linearly varying
with the QBER $q_1$, as synthesized by Eqs.~\eqref{q1_IR} and
Fig.~\ref{fig:IRinfos}. It is possible for Eve to increase her
information as a function of $q_1$ by performing non-orthogonal
measurements~\cite{Gisin02} on both the forward and the backward
paths. This strategy has been initially described in~\cite{Luc05}
by considering the mutual information as function of the
``detection probability''. Now we use the same approach but in
terms of QBER.

Given the four states prepared by Bob, and Eve's ancillary states
$|\varepsilon\rangle$, we can write the most general operation Eve
can do on traveling qubit at point $E_{1}$ of
Fig.\ref{fig:1-protocol} as:
\begin{align}
\left|  0\right\rangle \left|  \varepsilon\right\rangle  &
\rightarrow\left| 0\right\rangle \left|
\varepsilon_{00}\right\rangle +\left|  1\right\rangle \left|
\varepsilon_{01}\right\rangle =\sqrt{F}\left|  0\right\rangle
\left| \widetilde{\varepsilon}_{00}\right\rangle +\sqrt{D}\left|
1\right\rangle
\left|  \widetilde{\varepsilon}_{01}\right\rangle, \nonumber\\
\left|  1\right\rangle \left|  \varepsilon\right\rangle  &
\rightarrow\left| 0\right\rangle \left|
\varepsilon_{10}\right\rangle +\left|  1\right\rangle \left|
\varepsilon_{11}\right\rangle =\sqrt{D}\left|  0\right\rangle
\left| \widetilde{\varepsilon}_{10}\right\rangle +\sqrt{F}\left|
1\right\rangle
\left|  \widetilde{\varepsilon}_{11}\right\rangle, \nonumber\\
\left|  +\right\rangle \left|  \varepsilon\right\rangle  &
\rightarrow \frac{1}{\sqrt{2}}\left[  \left|  0\right\rangle
\left(  \left| \varepsilon_{00}\right\rangle +\left|
\varepsilon_{10}\right\rangle \right) +\left|  1\right\rangle
\left(  \left|  \varepsilon_{01}\right\rangle +\left|
\varepsilon_{11}\right\rangle \right)  \right] \nonumber\\
&  \equiv\left|  +\right\rangle \left|
\varepsilon_{++}\right\rangle +\left| -\right\rangle \left|
\varepsilon_{+-}\right\rangle,
\nonumber\\
\left|  -\right\rangle \left|  \varepsilon\right\rangle  &
\rightarrow \frac{1}{\sqrt{2}}\left[  \left|  0\right\rangle
\left(  \left| \varepsilon_{00}\right\rangle -\left|
\varepsilon_{10}\right\rangle \right) +\left|  1\right\rangle
\left(  \left|  \varepsilon_{01}\right\rangle -\left|
\varepsilon_{11}\right\rangle \right)  \right] \nonumber\\
&  \equiv\left|  +\right\rangle \left|
\varepsilon_{-+}\right\rangle +\left| -\right\rangle \left|
\varepsilon_{--}\right\rangle, \label{Nonorthogonal_E1}
\end{align}
where the states $\left|  \varepsilon_{ij}\right\rangle $ belong
to the four-dimensional Hilbert space of Eve's probe. Ancillary
states with tilde are intended to be normalized. The following
conditions make the transformations (\ref{Nonorthogonal_E1})
unitary:
\begin{align}
\left\langle \varepsilon_{00}|\varepsilon_{00}\right\rangle
+\left\langle \varepsilon_{01}|\varepsilon_{01}\right\rangle  &
\equiv
F+D=1,\label{Unitarity Conditions_1}\\
\left\langle \varepsilon_{10}|\varepsilon_{10}\right\rangle
+\left\langle \varepsilon_{11}|\varepsilon_{11}\right\rangle  &
\equiv
D+F=1,\label{Unitarity Conditions_2}\\
\left\langle \varepsilon_{00}|\varepsilon_{10}\right\rangle
+\left\langle \varepsilon_{01}|\varepsilon_{11}\right\rangle  &
=0.\label{Unitarity Conditions_3}%
\end{align}
Within condition (\ref{Unitarity Conditions_3}) we can set
$\left\langle \varepsilon_{00}|\varepsilon_{10}\right\rangle
=\left\langle \varepsilon_{01}|\varepsilon_{11}\right\rangle =0$
{\tre if we assume that the attack brought about by Eve is
symmetric}~\cite{Gisin02}. {\tre A symmetric attack is obtained
when the disturbance introduced by Eve results to be the same in
the two bases measured by the users. Alice and Bob can test the
occurrence of a symmetric attack by measuring the noise levels in
the two bases separately and verifying that they are equal.
Intuitively, a symmetric attack corresponds to the symmetric
structure of the states prepared by the transmitter, both in the
LM05 and in the BB84 protocol. In fact, for the BB84, it has been
shown to be optimal (see Section VI.E of~\cite{Gisin02}). Our
results are then limited to this class of individual symmetric
attacks.} We specify the angles between non-orthogonal vectors as:
\begin{equation}
\left\langle \widetilde{\varepsilon}_{00}|\widetilde{\varepsilon}%
_{11}\right\rangle =\cos x\,,\qquad\left\langle \widetilde{\varepsilon}%
_{01}|\widetilde{\varepsilon}_{10}\right\rangle =\cos
y,\label{Nonorthogonality Condition}
\end{equation}
with $0\leq x\,,\,y\leq\pi/2$. In this sense the present strategy
represents a non-orthogonal attack. Here, we do not fix values of
the parameters we introduced.

At point $E_{2}$ of Fig.\ref{fig:1-protocol} Eve performs an
attack similar to that at point $E_{1}$, but with fresh ancillae
$|\eta\rangle$ (hence new parameters $F^{\prime}$ and $D^{\prime}$
are in order), to gain information about Alice's encoding:
\begin{align}
\left|  0\right\rangle \left|  \eta\right\rangle  &  \rightarrow
\sqrt{F^{\prime}}\left|  0\right\rangle \left|  \widetilde{\eta}
_{00}\right\rangle +\sqrt{D^{\prime}}\left|  1\right\rangle \left|
\widetilde{\eta}_{01}\right\rangle, \nonumber\\
\left|  1\right\rangle \left|  \eta\right\rangle  &  \rightarrow
\sqrt{D^{\prime}}\left|  0\right\rangle \left|  \widetilde{\eta}%
_{10}\right\rangle +\sqrt{F^{\prime}}\left|  1\right\rangle \left|
\widetilde{\eta}_{11}\right\rangle, \nonumber\\
\left|  +\right\rangle \left|  \eta\right\rangle  &
\rightarrow\left| +\right\rangle \left|  \eta_{++}\right\rangle
+\left|  -\right\rangle \left|
\eta_{+-}\right\rangle, \nonumber\\
\left|  -\right\rangle \left|  \eta\right\rangle  &
\rightarrow\left| +\right\rangle \left|  \eta_{-+}\right\rangle
+\left|  -\right\rangle \left|
\eta_{--}\right\rangle. \label{Nonorthogonal_E2}
\end{align}
At the end of the transmission Eve will measure $\varepsilon$ and
$\eta$ ancillae and she will gain information. Our next task is to recover the optimal eavesdropping
strategy by Eve, i.e. determine parameters' values that maximize
Alice-Eve and Bob-Eve mutual information ($I_{AE}$, $I_{BE}$)
minimizing the QBERs $q_1$ and $Q_{AB}$.

From transformations
(\ref{Nonorthogonal_E1}) and conditions
(\ref{Unitarity Conditions_1}-\ref{Unitarity Conditions_3}) we can
easily evaluate $q_1$ for each basis prepared by Bob:
\begin{eqnarray}
  q_1(\Z) &=& \left\langle \varepsilon_{01}
|\varepsilon_{01}\right\rangle =\left\langle \varepsilon_{10}
|\varepsilon_{10}\right\rangle =D\label{q1_Z}, \\
\nonumber  q_1(\X) &=& \left\langle \varepsilon_{+-}
|\varepsilon_{+-}\right\rangle = \left\langle \varepsilon_{-+}
|\varepsilon_{-+}\right\rangle \\
          &=&\left[  1-F\cos x-D\cos y\right]/2. \label{q1_X}
\end{eqnarray}

A good choice for Eve to minimize the QBER $q_1$ is to align her
measuring apparatus with $\Z$ and $\X$ at random. If she aligns
along $\Z$ then $q_1(\Z) = D = 0$, otherwise it will be $q_1(\X)$
to be zero. On average, after setting $D=0$ (and so $F=1$), we
will have:
\begin{equation}\label{eq:q1_av}
    q_1^{av}=\frac{1-\cos x}{4},
\end{equation}
and Eqs.(\ref{Nonorthogonal_E1})
become:
\begin{align}
\left|0\right\rangle \left|  \varepsilon\right\rangle  &
\rightarrow \left| 0\right\rangle \left|\varepsilon_{00}\right\rangle,\nonumber \\
\left|1\right\rangle \left|  \varepsilon\right\rangle  &
\rightarrow\left|  1\right\rangle \left|\varepsilon_{11}\right\rangle,\nonumber \\
\left|+\right\rangle \left|  \varepsilon\right\rangle  &
\rightarrow \frac{1}{\sqrt{2}}\left(  \left|  0\right\rangle
\left| \varepsilon_{00}\right\rangle +\left|  1\right\rangle
\left| \varepsilon_{11}\right\rangle   \right), \nonumber \\
\left|-\right\rangle \left|  \varepsilon\right\rangle  &
\rightarrow \frac{1}{\sqrt{2}}\left(  \left|  0\right\rangle
\left| \varepsilon_{00}\right\rangle -\left| 1\right\rangle \left|
\varepsilon_{11}\right\rangle  \right).
\end{align}

\noindent From the above equations it is clear that the only
relevant parameter is $x$: if it is zero then Eve's ancillae are
parallel and do not provide any information to her, nor do they
cause any error on the channel since any evolved state remains
equal to the initial one. On the contrary, if $x=\pi/2$ Eve's
ancillae are maximally informative to Eve, but maximum is also the
noise created on the channel, according to Eq.\eqref{eq:q1_av}.
Similar arguments hold for the backward path, after
$E_{2}$-attack, with primed parameters replacing not-primed ones.

In order to evaluate the mutual information $I_{AE}$ let us write Bob's initial
states as:
\begin{equation}
\left|  \Psi\right\rangle =\sum_{\alpha=0,1}c_{\alpha}\left|
\alpha \right\rangle, \label{Bob's initial states}
\end{equation}
where we made the following ansatzes correspondingly to the
initial states:
\begin{eqnarray}
\left|  0\right\rangle  &  \rightarrow & c_{\alpha}=\delta_{\alpha,0},\nonumber\\
\left|  1\right\rangle  &  \rightarrow & c_{\alpha}=\delta_{\alpha,1},\nonumber\\
\left|  +\right\rangle  &  \rightarrow & c_{\alpha}=\frac{1}{\sqrt{2}},\nonumber\\
\left|  -\right\rangle  &  \rightarrow & c_{\alpha}=\left(
-1\right)  ^{\alpha}
\frac{1}{\sqrt{2}}.
\end{eqnarray}
Then we can rewrite transformations (\ref{Nonorthogonal_E1}) as
\begin{eqnarray}
\left|  \Psi\right\rangle \left|  \varepsilon\right\rangle
=\sum_{\alpha =0,1}c_{\alpha}\left|  \alpha\right\rangle \left|
\varepsilon\right\rangle
\rightarrow\sum_{\alpha}c_{\alpha}\sum_{\beta}\left|
\varepsilon_{\alpha \beta}\right\rangle \left|  \beta\right\rangle,
\end{eqnarray}
and transformations
(\ref{Nonorthogonal_E2}) as
\begin{eqnarray}
\left|  \Psi\right\rangle \left|  \eta\right\rangle =\sum_{\alpha
=0,1}c_{\alpha}\left|  \alpha\right\rangle \left|
\eta\right\rangle
\rightarrow\sum_{\alpha}c_{\alpha}\sum_{\beta}\left|
\eta_{\alpha\beta }\right\rangle \left|  \beta\right\rangle.
\end{eqnarray}

Now suppose that Alice performs the identity $\I$ between the two
Eve's attacks. The following sequence describes the state
transformations:
\begin{eqnarray}
\left|  \Psi\right\rangle \left|  \varepsilon\right\rangle \left|
\eta\right\rangle
&&\overset{E_{1}}{\rightarrow}\sum_{\alpha}c_{\alpha}
\sum_{\beta}\left|  \beta\right\rangle \left|
\varepsilon_{\alpha\beta
}\right\rangle \left|  \eta\right\rangle \nonumber\\
&&\overset{I}{\rightarrow}   \sum_{\alpha}c_{\alpha}
\sum_{\beta}\left|  \beta\right\rangle \left|
\varepsilon_{\alpha\beta }\right\rangle \left|
\eta\right\rangle \nonumber \\
&&\overset{E_{2}}{\rightarrow}
\sum_{\alpha}c_{\alpha}\sum_{\beta,\gamma}\left|
\gamma\right\rangle \left| \varepsilon_{\alpha\beta}\right\rangle
\left|  \eta_{\beta\gamma}\right\rangle.\label{Identity}
\end{eqnarray}
The ancillary states involved in this operation are:
\begin{align}
&  \left|  \varepsilon_{00},\eta_{00}\right\rangle ,\left|
\varepsilon _{00},\eta_{01}\right\rangle ,\left|
\varepsilon_{01},\eta_{10}\right\rangle
,\left|  \varepsilon_{01},\eta_{11}\right\rangle, \nonumber\\
&  \left|  \varepsilon_{10},\eta_{00}\right\rangle ,\left|
\varepsilon _{10},\eta_{01}\right\rangle ,\left|
\varepsilon_{11},\eta_{10}\right\rangle ,\left|
\varepsilon_{11},\eta_{11}\right\rangle
.\label{AncillaryStatesInIdentity}%
\end{align}

If, instead of  the identity $\I$, Alice performs the $i\Y$
operation we have:
\begin{eqnarray}
\left|  \Psi\right\rangle \left|  \varepsilon\right\rangle \left|
\eta\right\rangle
&&\overset{E_{1}}{\rightarrow}\sum_{\alpha}c_{\alpha}
\sum_{\beta}\left|  \beta\right\rangle \left|
\varepsilon_{\alpha\beta
}\right\rangle \left|  \eta\right\rangle \nonumber\\
&&\overset{iY}{\rightarrow}
\sum_{\alpha}c_{\alpha}\sum_{\beta}\left( -1\right)
^{\beta+1}\left|  \beta\oplus1\right\rangle \left|  \varepsilon
_{\alpha\beta}\right\rangle \left|  \eta\right\rangle \nonumber\\
&&\overset{E_{2}}{\rightarrow}
\sum_{\alpha}c_{\alpha}\sum_{\beta,\gamma }\left(  -1\right)
^{\beta+1}\left|  \gamma\right\rangle \left|
\varepsilon_{\alpha\beta}\right\rangle \left|  \eta_{\left(  \beta
\oplus1\right)  \gamma}\right\rangle ,\label{Flip}%
\end{eqnarray}
and the involved ancillary states are:
\begin{align}
&  \left|  \varepsilon_{00},\eta_{10}\right\rangle ,\left|
\varepsilon _{00},\eta_{11}\right\rangle ,\left|
\varepsilon_{01},\eta_{00}\right\rangle
,\left|  \varepsilon_{01},\eta_{01}\right\rangle, \nonumber\\
&  \left|  \varepsilon_{10},\eta_{10}\right\rangle ,\left|
\varepsilon _{10},\eta_{11}\right\rangle ,\left|
\varepsilon_{11},\eta_{00}\right\rangle ,\left|
\varepsilon_{11},\eta_{01}\right\rangle
.\label{AncillaryStatesInFlip}
\end{align}

In order to acquire information from states (\ref{Identity}) and
(\ref{Flip} ) Eve must measure both her ancillae. Keeping in mind
orthogonality relations (\ref{Unitarity Conditions_3}) and
following, we see that the best way to do that is to distinguish
orthogonal subspaces before, and then non-orthogonal states.
The probability to make an error in distinguishing between two
states with scalar product $\cos x$ is $\left(  1-\sin x\right) /2
$~\cite{Per97}. Observing states (\ref{Identity}) and (\ref{Flip})
we can notice that if Eve mistakes to identify her first ancilla
($\varepsilon$ states) then she guesses the wrong Alice's
operation, since she flips from states (\ref{Identity}) and
(\ref{Flip}) or viceversa. The same is true if she guesses right
$\varepsilon$ state but mistakes $\eta$ state. Nevertheless, if
she mistakes twice, then with the first error she misinterprets
(\ref{Identity}) and (\ref{Flip}) and with the second error she
compensates the first one, eventually guessing right Alice's
operation. This leads to estimate the probability Eve wrongly
guesses Alice's operation, i.e. the QBER $Q_{AE}$ between Alice
and Eve, and from it, considering $F=F'=1$, we find the following
expression for $I_{AE}$:
\begin{align}
I_{AE} &  =\left\{  1-H\left[  \left( \frac{1+\sin x}{2}
\right)  \left(  \frac{1+\sin x^{\prime}}{2}\right)  +\left(
\frac{1-\sin x}{2} \right)  \left( \frac{1-\sin
x^{\prime}}{2}\right)  \right]  \right\}.\label{Iae}
\end{align}
Then, to upper bound Eve's information without affecting $q_1$, we
can simply use orthogonal ancillae on the backward path, that
means to set $x^{\prime}=\pi/2$. In this case we obtain:
\begin{equation}
I_{AE}=1-H\left(  \frac{1-\sin x}{2}\right).
\label{Nonorthogonal_Info_AE}
\end{equation}

The last question concerns the mutual information between Bob and
Eve. This is limited by the fact that Bob does not reveal any
basis in LM05. So if Eve guesses the right basis in
transformations (\ref{Nonorthogonal_E1}) (probability equal to
1/2) then she has a probability of error equal to her capability
to discriminate her ancillae, given then by $(1-\sin x)/2$.
Otherwise, with probability 1/2, she guesses the wrong basis and
she has a probability of 1/2 to be wrong with the qubit also. In
the whole:
\begin{equation}
Q_{BE}=\frac{2-\sin x}{4}.\label{Q_BE}
\end{equation}

The security of the protocol now depends on the relation between
$Q_{AB}$ and $q_1$.
For the purpose of comparison we assume here the same model as per
Eq.\eqref{q1qab}, obtained for IR attacks: $Q_{AB}=q_1$. This is
reasonable as the present strategy is a generalization of IR to
nonorthogonal Eve's ancillae. With such an assumption it is
possible to draw the curves in Fig.~\ref{fig:NO-infos}.
\begin{figure}[h!]
\includegraphics[width=12.0cm]{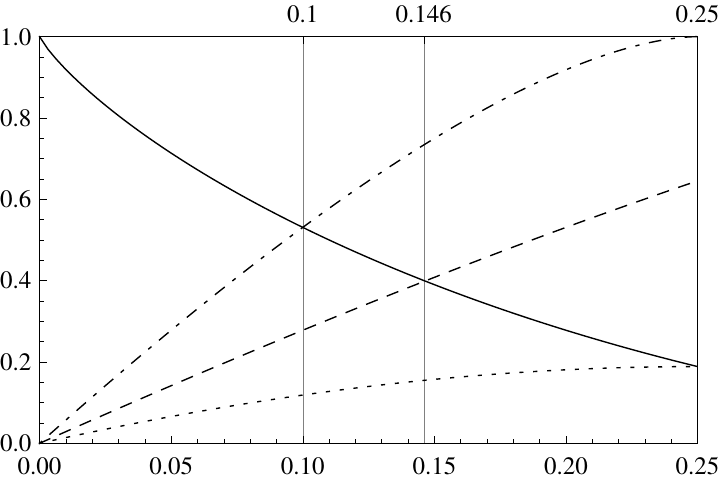}
\caption{Mutual information vs $q_1$ in NORT attack. The curves
are: $I_{AE}^{LM05}$ (dot-dashed line), $I_{BE}^{LM05}$ (dotted
line), $I_{AE}^{BB84} = I_{BE}^{BB84}$ (dashed line). The security
thresholds are reported at the top of the frame.}
\label{fig:NO-infos}
\end{figure}

\medskip

{\tre The increased information acquired by Eve using NORT rather
than IR is apparent. The security threshold for BB84 is now
$14.6\%$, equal to the one obtained in~\cite{FGN+97} for the
optimal single-particle eavesdropping. On the contrary, the
thresholds for LM05 are $10.0\%$ and $25.0\%$ for DR and RR
respectively. Hence LM05 used in RR is more tolerant to noise than
BB84, provided that Eve's action is limited to a NORT attack.}

\bigskip

\subsection{Double CNOT attack - DCNOT}
\label{subsec:DCNOT}

So far we have analyzed attacks in which Eve measures the qubit
coming out from Bob's station. As we have seen this kind of attack
can provide Eve with a high mutual information with Alice. However
the mutual information between Eve and Bob is poor, as illustrated
by the Figures~\ref{fig:IRinfos} and~\ref{fig:NO-infos}. The
reason is that Eve's initial measurement modifies the state
prepared by Bob, and prevents her from knowing the outcome of
Bob's final measurement.
%
%
It is then natural to ask whether is possible for Eve to perform
her attack without altering Bob's state, thus maximizing her
mutual information with Bob. This attack exists and has been
introduced for the first time in~\cite{LucPhd05}. It is composed
by a sequence of two CNOT gates from Eve, one for each path of the
LM05; hence we call it \textit{Double CNOT} attack, in short
``DCNOT''. We will show that DCNOT can make $I_{BE}$, distilled in
the RR modality, equal to its maximum and to $I_{AE}$, thus
reducing the LM05 security to the security obtained in the DR
modality.

Let us study the evolution of the states prepared by Bob under the
action of the first CNOT gate by Eve. Eve appends an ancilla in
the state $\left|  0\right\rangle _{e}$ to the initial states
prepared by Bob; then she performs a first CNOT gate before
Alice's station using the traveling qubit as control and her
ancilla as target. The states become:
\begin{align}
& \left|  0\right\rangle \left|  0\right\rangle _{e}\rightarrow
\left|  0\right\rangle \left|  0\right\rangle
_{e},\nonumber\\
& \left|  1\right\rangle \left|  0\right\rangle
_{e}\rightarrow\left|  1\right\rangle \left|  1\right\rangle
_{e},\nonumber\\
& \left|  +\right\rangle \left|  0\right\rangle
_{e}\rightarrow\frac{\left|  0\right\rangle \left| 0\right\rangle
_{e}+\left|  1\right\rangle \left| 1\right\rangle _{e}}{\sqrt{2}
},\nonumber\\
& \left|  -\right\rangle \left|  0\right\rangle
_{e}\rightarrow\frac{\left|  0\right\rangle \left| 0\right\rangle
_{e}-\left|  1\right\rangle \left| 1\right\rangle _{e}}{\sqrt{2}
}.\label{First CNOT}%
\end{align}
We can notice that when Bob prepares states in the basis $\X$ the
CNOT gate creates an entangled state with the traveling qubit and
Eve's ancilla. After that Eve forwards the  control qubit to
Alice, who performs her encoding $\hat{A}_{i}$ ($\hat{A}_0=\I$ and
$\hat{A}_1=i\Y$) on it and then sends it back to Bob. Eve, on the
backward path, executes a second CNOT gate on the whole system. We
report Alice's and Eve's actions as:
\begin{align}
& \left( \hat{A}_{i}\left|  0\right\rangle \right)  \left|
0\right\rangle_{e}\rightarrow\left(  \hat{A} _{i}\left|
0\right\rangle \right)  \left|  i\right\rangle_{e},\nonumber\\
& \left( \hat{A}_{i}\left|  1\right\rangle \right)  \left|
1\right\rangle_{e}\rightarrow\left(  \hat{A} _{i}\left|
1\right\rangle \right)  \left|  i\right\rangle_{e},\nonumber\\
& \frac{\left( \hat{A}_{i}\left|  0\right\rangle \right) \left|
0\right\rangle _{e}+\left(  \hat{A}_{i}\left| 1\right\rangle
\right) \left| 1\right\rangle_{e}}{\sqrt{2}}\rightarrow \left(
\hat{A}_{i}\left|
+\right\rangle \right)  \left|  i\right\rangle_{e},\nonumber\\
& \frac{\left( \hat{A}_{i}\left|  0\right\rangle \right) \left|
0\right\rangle_{e}-\left(  \hat{A}_{i}\left| 1\right\rangle
\right) \left| 1\right\rangle_{e}}{\sqrt{2}}\rightarrow -\left(
\hat{A}_{i}\left| -\right\rangle \right)  \left|
i\right\rangle_{e}.\label{Second CNOT}
\end{align}
From these equations we can recognize that the entanglement
created by Eve in the first CNOT gate disappears with the second
CNOT. Furthermore, Eve's ancillae take exactly the information
encoded by Alice; hence it is sufficient for Eve to measure them
in the basis $\Z$ to find out Alice's encoding: $\left|
0\right\rangle _{e}$ indicates that Alice performed the identity,
while $\left| 1\right\rangle _{e}$ indicates the spin-flip
operation. Finally, after the second CNOT the state arriving to
Bob is just the right state he expected to receive! This entails
two consequences:

i) in the DCNOT $Q_{AB}=0$. If Alice and Bob use only $Q_{AB}$ as
a security parameter, they will never detect this kind of attack
by Eve. However, for the same reason, the mutual information
between Alice and Bob, which is a function of $Q_{AB}$, is always
equal to 1.

ii) Eve acquires full information about both Alice ($I_{AE}$) and
Bob ($I_{BE}$). In fact she knows perfectly the encoded
transformation and the result that is going to be obtained by Bob.

The DCNOT attack can be detected in CM. If the qubit is prepared
in the basis $\Z$ Eve does not perturb the state at all while if
it was prepared in the basis $\X$ the QBER $q_1$ is equal to 1/2.
Hence the overall QBERs and information situation for this attack
is:
\begin{eqnarray}
q_1 &=& 0.25, \\
\label{eq:QBER_CNOT1}
Q_{AB} &=& 0 \Rightarrow I_{AB}=1, \\
\label{eq:QBER_CNOT2}
I_{AE} &=& I_{BE} = 1.
\label{eq:QBER_CNOT3}
\end{eqnarray}
These equations are quite similar to Eqs.~\eqref{q1_IR} obtained
for the IR attack, with the important difference that $Q_{AB}$ is
now zero {\due and, by consequence, $I_{AB}$ is always equal to 1,
as per Eq.\eqref{eq:QBER_CNOT2}. Hence, by making use of the CK
theorem, Eq.\eqref{CKtheorem}, we can conclude that the LM05
provides a secure rate against DCNOT over the whole interval of
$q_1$}.

Likewise IR, if Eve performs the DCNOT in a partial way, her
information will be a linear function of the QBER $q_1$; then the
corresponding curve would be exactly the same as the one depicting
$I_{AE}^{LM05}$ in Fig.~\ref{fig:IRinfos}. {\gen We report such a
curve in Fig.~\ref{fig:DCNOT}}.

{\gen An important modification to the DCNOT is the following. Eve
executes a random flip/no-flip unitary operation, analogous to the
one effected by Alice, after the second CNOT, on the backward
path. In this way Eve increases $Q_{AB}$, from its minimum value
$0$ to any desired value $\chi$ and decreases the mutual
information between Alice and Bob from its maximum value $1$ to
any other value imposed by her. Furthermore, with this kind of
modified DCNOT attack, which we term DCNOT$^*$, Eve still
maintains the control about Bob's final measurement outcome, as in
the non-modified DCNOT.}
\begin{figure}[h!]
  \includegraphics[width=12.0cm]{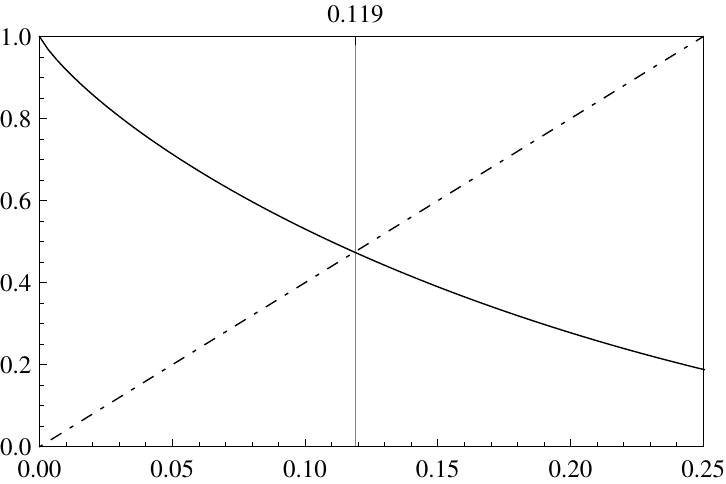}\\
  \caption{Mutual information $I_{AB}$ (solid line) and
  $I_{AE}^{LM05}=I_{BE}^{LM05}$ (dot-dashed line) versus $q_1$ in the
  DCNOT attack against LM05. The security threshold is reported at the top of the frame.}
  \label{fig:DCNOT}
\end{figure}
The situation for the DCNOT$^*$ attack is then summarized by the
following equations:
\begin{eqnarray}\label{eq:QBER_CNOT_2}
\nonumber  q_1 &=& 0.25 \xi, \\
\nonumber  Q_{AB} &=& \chi \Rightarrow I_{AB}=1-H(\chi), \\
  I_{AE} &=& I_{BE} = \xi,
\end{eqnarray}
with $\chi\in[0,0.5]$ a parameter controlled by Eve and $\xi$ the
fraction of attacked qubits, as already defined for IR attacks.
{\gen We can again assume the noise model of Eq.\eqref{q1qab} to
compare this new attack with the previous ones, i.e. $Q_{AB}=q_1$.
The corresponding curves are plotted in Fig.~\ref{fig:DCNOT}.

The LM05 protocol results secure, both in DR and in RR, if
$q_1<11.9\%$. Hence, by comparing with Figs.~\ref{fig:IRinfos}
and~\ref{fig:NO-infos}, it is seen that this attack is less
dangerous than NORT as far as $I_{AE}$ is concerned, but is far
more dangerous for what concerns the security threshold pertaining
to $I_{BE}$, which decreases from $25.0\%$ to $11.9\%$.}

\subsection{Generic individual attack} \label{sec:GenSec}

%
%
{\tre In this section we provide an upper bound to Eve's
information in case of single-particle attack. We adapt to LM05 an
argument recently introduced in~\cite{CL04}, which in turn is
based on the main argument of~\cite{LC99} limited to the case of
an individual attack. We remark that such a proof holds when LM05
is used for QKD, not for a deterministic Direct
Communication~\cite{BF02}. We also stress that the proof given
in~\cite{CL04} and repeated here for LM05 does not represent a
security proof against the most general attack by Eve, i.e. a
coherent attack of a two-way deterministic protocol. In fact, the
argument does not include any multi-particle distillation of
quantum states performed by the users, nor multi-particle attacks
by Eve~\cite{NoteUncSec}. }

{\tre
As a first step of the generic individual attack by Eve suppose
that the users are provided by Eve with a state which is claimed
to be a perfect singlet:
\begin{equation}
\left\vert \psi_{AB}\right\rangle =\left( \left\vert
0\right\rangle _{A}\left\vert 1\right\rangle _{B}-\left\vert
1\right\rangle _{A}\left\vert 0\right\rangle _{B}\right)
/\sqrt{2}, \label{singlet}
\end{equation}
where the states $\left\vert 0\right\rangle$ and $\left\vert
1\right\rangle$ are the eigenstates of the basis $Z$, the
computational basis.
If the state~\eqref{singlet} were a truly singlet, the users would
perform LM05 as follows. Bob measures his particle in one of the
two bases $Z$ or $X$ chosen at random, thus preparing Alice's
particle's state in a state orthogonal to his state. In EM, Alice
performs on her particle the desired operation $\one$ or $iY$,
exactly as she would do in the standard prepare-and-measure LM05.
All the same, in CM, Alice measures her particle in a basis
randomly chosen between $Z$ or $X$. This reduces the
entanglement-based LM05 to a prepare-and-measure protocol. Once
the security is shown for the former it is automatically true for
the latter. }

{\tre The next task is for the users to verify that they actually
have been given by Eve the claimed state. In practice they must
verify that the fidelity $F(\left\vert \psi_{AB}\right\rangle,
\rho_{AB})$ between the claimed state and the state $\rho_{AB}$ in
their hands is very close to 1. Hence consider a tripartite state
$\left\vert \Phi\right\rangle _{ABE}$ which is a purification of
$\rho_{AB}$ and which is in Eve's hands:
\begin{eqnarray}
  \rho_{AB} &=& tr_{E}\left(  \left\vert \Phi\right\rangle
_{ABE}\left\langle \Phi\right\vert \right) \\
  \rho_{E} &=& tr_{AB}\left( \left\vert
\Phi\right\rangle _{ABE}\left\langle \Phi\right\vert \right).
\end{eqnarray}
Then $S(\rho_{AB})=S\left( \rho_{E}\right) $, where $S$ is the Von
Neumann entropy~\cite{NC00}, because $\left\vert \Phi\right\rangle
_{ABE}$ is pure and then $\rho_{AB}$ and $\rho_{E}$ have the same
spectrum. Hence Alice and Bob can calculate Eve's maximum
information from the relation $S(\rho_{AB})=S\left(
\rho_{E}\right)$ provided they can estimate $S(\rho_{AB})$. To
this aim the users perform the CM. In particular they measure the
QBER $q_1$ which is equal to the \textit{infidelity} $\delta$ of
Ref.~\cite{LC99}, which is defined by the relation
\begin{equation}\label{fidelitybound}
    F\left(\left\vert \psi_{AB}\right\rangle,
\rho_{AB}\right)^{2}\geq1-\delta.
\end{equation}
As a result, if $q_1$ is not too large, the state $\rho_{AB}$ is
acknowledged to be very similar to $\left\vert
\psi_{AB}\right\rangle $ by the users. In particular, since they
verify that $F\left(\left\vert \psi_{AB}\right\rangle,
\rho_{AB}\right)^{2} \geq1-q_{1}$ then from the Lemma ``High
fidelity implies low entropy'' of Ref.~\cite{LC99} one has:
\begin{equation}
S(\rho_{AB})=S\left(  \rho_{E}\right) \leq -\left(  1-q_{1}\right)
\log _{2}\left(  1-q_{1}\right)  -q_{1}\log_{2}\left(
\frac{q_{1}}{3}\right).
\label{e1}
\end{equation}
This represents an upper bound to Eve's absolute information at
the end of the forward path, in full analogy with what happens in
the BB84 protocol. What we need now is a connection with the
backward path. Between the two paths, Alice encodes information on
her particle. So let us make a step back and focus on the mutual
information between Alice and Eve, $I_{AE}$.

It is known that the \textit{Holevo information} is un upper bound
to the mutual information~\cite{NC00}:
\begin{equation}\label{Hol1}
    I_{AE} \leq \chi(\rho_{AE}),
\end{equation}
with $\rho_{AE}=tr_{B}\left(\left\vert
\Phi\right\rangle_{ABE}\left\langle \Phi\right\vert \right)$ and
$\chi(\rho)=S(\rho)-\sum_k S(\rho_k)$. Furthermore the following
relations trivially hold:
\begin{equation}\label{Hol2}
    \chi(\rho_{AE}) \leq S(\rho_{AE}) \leq S(\rho_E),
\end{equation}
with $\rho_{E}=tr_{A}\left(\rho_{AE}\right) $. The last equation
is an intuitive corollary to the Holevo theorem cited as ``Lemma
2'' in Ref.~\cite{LC99}. So, it turns out that the upper bound
$S(\rho_E)$ available to Alice and Bob is an upper bound both to
the mutual information $I_{AE}$ and to the Holevo quantity
$\chi(\rho_{AE})$. As stated in Ref.~\cite{CL04} the Holevo
quantity does not increase under quantum operations. In particular
Alice's encoding operations cannot increase $\chi(\rho_{AE})$. By
consequence one has:
\begin{equation}\label{S-bef-aft}
    \chi(\rho_{AE})^{after} \leq \chi(\rho_{AE})^{before} \leq S(\rho_E)
\end{equation}
where the labels ``before'' and ``after'' refer to Alice's
encoding. So $S(\rho_E)$ represents a bound to Eve's information
even after Alice's encoding.

To conclude the proof it suffices to note that even if Eve can
increase $S(\rho_E)$ on the backward path, e.g. by attaching new
ancillae to the traveling states, the increase in information does
not concern Alice's encoding. So it would be the same for Eve if
she increases her information since the beginning on the forward
path; but in that case the users already have the bound of
Eq.\eqref{S-bef-aft}.

Now it is possible to use the CK theorem, Eq.\eqref{CKtheorem},
and the noise model of Eq.\eqref{q1qab} to make a guess about how
tight the given bound is. In this model the mutual information
$I_{AB}$ is given by $1-H(q_1)$ while from Eq.\eqref{e1} we get un
upper bound to Eve's average information.
The curves pertaining to the mutual information are plotted in
Fig.~\ref{fig:GenSec}. They are compatible with the previous
results since the found threshold, $q_1\leq8.8\%$, is lower than
all previous single-particle attacks, both in DR and in RR.
\begin{figure}
  \includegraphics[width=12.0cm]{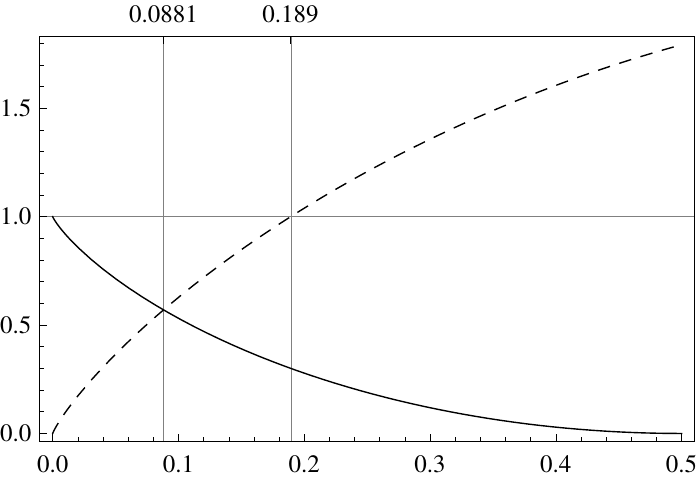}\\
  \caption{General attack security for LM05 in DR. The security
  threshold amounts to $q_1=8.8\%$. For $q_1>18.9\%$ Eve gets full information
  about Alice's encoding.}
  \label{fig:GenSec}
\end{figure}
Note however that the given bound is meaningful only for LM05 in
DR since it is based on the Holevo information bound
Eq.\eqref{S-bef-aft} which makes explicit reference to Alice's
encoding and none to Bob's measurement. However although the
security threshold pertaining to RR could be higher than that for
DR, it cannot be too much different because we provided an
explicit attack, the DCNOT$^*$, which gives a security threshold
of $11.9\%$ for LM05 in RR, only $35\%$ higher than the above
threshold. Moreover, limiting the analysis to DR, it turns out
that the security bound just given is quite tight. In fact, with
the NORT attack we found a security threshold of $10.0\%$, which
is only $13.6\%$ higher than the given bound.

Finally let us note that the above bound holds trivially for BB84
as well, where Alice does not encode information on the received
state. In this case however the bound is not tight as $8.8\%$ is
far smaller than the $14.6\%$ pertaining to the optimal individual
attack against BB84~\cite{FGN+97}.}
}

{\tre Before moving to the next Section and study the zero-QBER
attacks, we summarize the results found for the zero-loss
individual eavesdropping in Table~\ref{Table2}.
\begin{table}[h!]
  \centering
  \begin{tabular} [c]{|c|c|c|c|}\hline
Eve's strategy & LM05-DR (\%) & LM05-RR (\%) & BB84 (\%)\\
\hline $IR$ & \multicolumn{1}{|r|}{11.9} &
\multicolumn{1}{|r|}{25.0} & \multicolumn{1}{|r|}{17.1}\\
\hline $NORT$ & \multicolumn{1}{|r|}{10.0} &
\multicolumn{1}{|r|}{25.0} & \multicolumn{1}{|r|}{14.6}\\
\hline $DCNOT^*$ & \multicolumn{1}{|r|}{11.9} &
\multicolumn{1}{|r|}{11.9} & \multicolumn{1}{|r|}{x}\\
\hline $Generic$ & \multicolumn{1}{|r|}{8.8} &
\multicolumn{1}{|r|}{x} & \multicolumn{1}{|r|}{8.8}\\
\hline
\end{tabular}
  \caption{security thresholds for single-particle
attacks against LM05 in DR, LM05 in RR and BB84.}\label{Table2}
\end{table}
}

\section{Zero-QBER eavesdropping}
\label{sec:0QBER}

We now consider attacks that introduce losses, but no noise. If
Alice uses a perfect single-photon source the single photon can
only be detected \textit{either} by Eve \textit{or} by Bob.
However in practice Alice uses an approximated single-photon
source i.e. an attenuated laser beam. For this source the number
of photon per pulse follows a Poisson statistics. It is then
possible that a single pulse contains more than one photon in the
same quantum state, and that \textit{both} Eve \textit{and} Bob
detect the same qubit. In this case Eve steals one bit of
information without introducing any noise in Bob's measurement;
however she introduces losses. Alice and Bob, from the knowledge
of the loss rate, should be able to infer how much information has
been stolen and remove it by means of Privacy
Amplification~\cite{Bennett1988,Ben95pa}. It is worth noting that
in the hypothesis of zero-QBER attacks the mutual information
between Alice and Bob is always maximum and no Error Correction is
needed to reconcile users' keys. Another obvious consequence of
this assumption is that Eve does not acquire any further
information from the Error Correction procedure. So her knowledge
will derive entirely from her attack to the quantum channel.


\subsection{Beam splitting attack - BS}
\label{subsec:BS}

In the Beam Splitting attack (BS) Eve uses a beam-splitter of
transmissivity $T=1-R$ to deviate a fraction $R$ of the main beam
towards her detectors, which are supposed to be ideal (100\%
quantum efficiency, no dark counts). The approach for BS is very
similar to the one adopted for the IR attack: we first evaluate
the probability of success for an Eve eavesdropping one bit of
information; then we relate the photon emitted by the source to
the eavesdropped bits via the Binomial distribution.

Let us first analyze the BB84~\cite{Bennett92}.
%
%
Define $\mu$ as the average number of photons contained in each
pulse prepared by Alice. With her beam-splitter Eve splits a
fraction $R\mu$ from the main beam towards her (ideal) detectors,
so that the probability to successfully detect a photon is given
by:
\begin{equation}
P=1-e^{-R\mu}\approx R\mu.
\end{equation}
The above approximation holds when $\mu\ll1$. Usually in the
experiments on BB84 $\mu=0.1$. It is reasonable to assume that
$R\simeq1$ because usually the attenuation rate measured by Alice
and Bob is very close to the unity at wavelength of 1550 nm. Hence
Eve's probability to successfully detect a photon is about $\mu$.
We also conservatively assume that Eve possesses a perfect quantum
memory to store the bits until the moment in which the basis is
revealed by Alice and Bob, so that every stored photon is detected
by Eve in the right basis. This means that $\mu$ represents the
average information collected by Eve against BB84 through the BS
attack in the asymptotic limit of a large number $N$ of pulses
traveling on the channel:
\begin{equation}\label{eq:BS-BB84}
I_{E}^{\textrm{\textsc{BB84}}} =\mu.
\end{equation}
{\gen This expression can be used to provide a secure gain similar
to that obtained for IR attack. Specifically the secure gain
$G^{BB84}$ is:
\begin{equation}\label{eq:secureGainBB84}
G^{BB84}=G_{raw}^{BB84}\times\left( 1-I_{E}^{BB84} \right),
\end{equation}
where:
\begin{align}
G_{raw}^{BB84}  &  =1-\exp\left[
-\mu\times\eta_{d}\times\Gamma^{BB84}\left(
L\right)  \right] \label{eq:GrawBB84}, \\
\Gamma^{BB84}\left(  L\right)   &  =\Gamma_{QC}\left(  L\right)
\times
\Gamma_{B}\text{ \ (total transmission)},\\
\Gamma_{QC}\left(  L\right)   &  =10^{-0.02\times L}\text{
(transmission of the quantum channel)},
\end{align}
with parameters $\eta_{d}=0.12$ (efficiency of Bob's detectors),
$\Gamma_{B}=0.4$ (transmission of Bob's box), and $\mu$ is
optimized for every distance $L$ in order to give the maximum
secure gain.
}

In analogy with BB84 one can calculate the amount of Privacy
Amplification~\cite{Bennett1988,Ben95pa} necessary to cope with BS
in LM05. Being LM05 a two-way protocol the BS must be accomplished
using two beam-splitters rather than one, positioned in the two
paths of the communication channel. In this way it may happen that
Eve measures two photons, one from the forward path and one from
the backward one; then, by comparing her outcomes, Eve can
ascertain the encoded information. Let $R_{1}$ and $R_{2}$ be the
reflectivity of the two beam-splitters. After the first
beam-splitter Eve has a probability of successful detection
\begin{equation}
P_{1}=1-e^{-R_{1}\mu}.
\end{equation}
The fraction of the beam transmitted through the first
beam-splitter is $T_{1}\mu=(1-R_{1})\mu$. Then the fraction of the
beam deviated by the second beam-splitter is $R_{2}(1-R_{1})\mu$,
and Eve's probability to detect a photon in the second path is:
\begin{equation}
P_{2}=1-e^{-R_{2}(1-R_{1})\mu}.
\end{equation}
A successful eavesdropping is given by two successful detection
events in the same run:
\begin{equation}
P_{BS_{12}}=P_{1}\cdot P_{2}=\left(  1-e^{-R_{1}\mu}\right) \left(
1-e^{-R_{2}(1-R_{1})\mu}\right)  . \label{P_TOT}%
\end{equation}
From this expression it is straightforward to check that the value
of $R_{1}=1/2$ maximizes $P_{BS_{12}}$ for any $\mu$. An intuitive
way to understand this is to take $R_{2}\simeq1$, to let Eve read
as much information as possible from the backward path, and to
realize that in this case Eve's best choice is to analyze half of
the beam from the forward path and half from the backward one,
i.e. to set $R_{1}=1/2$. Inserting these values into
Eq.(\ref{P_TOT}) we have:
\begin{equation}
P_{BS_{12}}\leq\left(  1-e^{-\mu/2}\right)  ^{2}\equiv
P^{\ast}=1-2\allowbreak e^{-\mu/2}+\allowbreak e^{-\mu},
\end{equation}
which represents the average information acquired by Eve through
the BS attack against the LM05 protocol, both in DR and RR, in the
asymptotic limit of an infinite number of pulses traveling on the
two-way quantum channel~\cite{Kum08}:
\begin{align}
\label{eq:BS-LM05} I_{E}^{\textrm{\textsc{LM05}}} = P^{\ast}.
\end{align}
{\gen This expression can be used to provide a secure gain similar
to that previously obtained for BB84. Specifically the secure gain
$G^{LM05}$ is:
\begin{equation}\label{eq:secureGainLM05}
G^{LM05}=G_{raw}^{LM05}\times\left( 1-I_{E}^{LM05} \right),
\end{equation}
where:
\begin{align}
G_{raw}^{LM05}  &  =1-\exp\left[
-\mu\times\eta_{d}\times\Gamma^{LM05}\left(
L\right)  \right] \label{eq:GrawLM05}, \\
\Gamma^{LM05}\left(  L\right)   &  =\Gamma_{QC}^{2}\left( L\right)
\times\Gamma_{B}\times\Gamma_{A}^{2}\text{ \ (total transmission)},\label{eq:GrawLM05b}\\
\Gamma_{QC}\left(  L\right)   &  =10^{-0.02\times L}\text{
(transmission of the quantum channel),}\label{eq:GrawLM05b}
\end{align}
with parameters $\eta_{d}=0.12$ (efficiency of Bob's detectors),
$\Gamma_{B}=0.4$ (transmission of Bob's box)}, $\Gamma_{A}=0.45$
(transmission of Alice's box) and $\mu$ is optimized for every
distance $L$.
Note that the crucial equation is that defining
$\Gamma^{LM05}\left( L\right)$ which, with respect to BB84,
contains the square of $\Gamma _{QC}\left(  L\right)$, because the
channel is two-way, and the term $\Gamma_{A}^{2}$, which
represents the double-transmission of Alice's setup because the
photon passes twice in it before going back to Bob. The values of
the parameters reported after Eq.\eqref{eq:GrawLM05b} are similar
to those experimentally measured in Ref.~\cite{Kum08}.
\begin{figure}[h!]
\includegraphics[width=12.0cm]{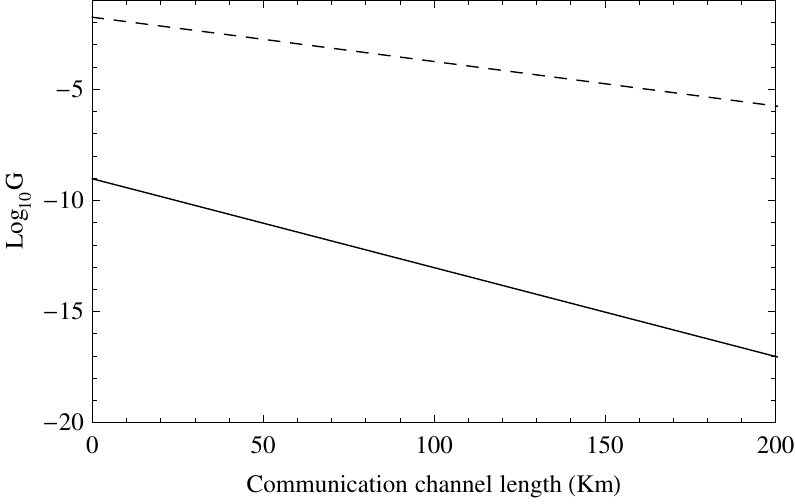}
\caption{Plot of the secure gain of BB84 ($G^{BB84}$,
Eq.\eqref{eq:secureGainBB84}, dashed line) and LM05 ($G^{LM05}$,
Eq.\eqref{eq:secureGainLM05}, solid line) versus the length of the
communication channel, in case of BS attack by Eve. The BB84 is
superior to LM05 for all the distances. This is due to the double
loss-rate of the two-way LM05 protocol.} \label{Fig.BS}
\end{figure}

In Figure~\ref{Fig.BS} we plot $\log_{10} G^{BB84}$ and $\log_{10}
G^{LM05}$. It can be noted that the secure rate pertaining to LM05
is far smaller than that pertaining to BB84. This crucially
depends on the higher loss-rate of the two-way channel of LM05.

However in the next Section we show that the difference between
the two protocols in the frame of zero-QBER attacks can be
dramatically reduced when the Photon-Number Splitting (PNS) attack
by Eve is considered. This is due to the higher resistance offered
by LM05 against this kind of attacks.

\subsection{Photon-number splitting attack - PNS}
\label{subsec:PNS}

In the following we describe the Photon-Number Splitting (PNS)
attack~\cite{HIG+95,Yue96,Lut00,Brassard2000,LJ02} against the
present version of the LM05 protocol. This attack has been
previously described in~\cite{Luc06} for the original LM05
protocol. The removal of the test on the backward channel does not
present any major subtleties: it is simply replaced by the test of
the QBER $Q_{AB}$, which obliges Eve to perform an attack on at
least three photons (see the following explanation) to not
introduce any disturbance on the channel.

As mentioned in the previous Section~\ref{subsec:BS}, when the
photon source is a laser attenuated with an average photon number
per pulse $\mu$, the probability to have $n$ photons in a single
pulse is given by a Poisson distribution:
\begin{equation}
P_{n}\left(  \mu\right)
=\frac{\mu^{n}}{n!}e^{-\mu}.\label{Pn(miu)}
\end{equation}
This means that with a probability $P_{n}\left(  \mu\right)  $\
Bob prepares the state $ \left\vert \psi\right\rangle ^{\otimes
n}$ rather than the desired state $\vert \psi \rangle $ ($\psi$
indicates one of the four states prepared by Bob in BB84 and
LM05). This accidental redundancy can be exploited by Eve to
perform a perfect zero-QBER attack.

It is known~\cite{SARG04} that when $n=3$ it exists a measurement
$\mathcal{M}$ that provides a conclusive result about the absolute
polarization $\psi$ with (optimal) probability $1/2$. Eve can
exploit this fact to eavesdrop on LM05 protocol in the following
way. She performs a quantum nondemolition measurement (QND) on the
pulses as soon as they exit Bob's station; this can be done
without perturbing the state $\psi$: when she finds $n<3$ she
blocks the pulses; on the pulses with at least three photons she
executes $\mathcal{M}$ and if the outcome is not conclusive she
blocks these pulses as well; when $n\geq3$ and the outcome\ of
$\mathcal{M}$ is conclusive she prepares a new photon in the right
state $\psi$ and forwards it to Alice. Until here this attack is
completely analogous to the `IRUD-attack' described in
\cite{SARG04}. The only variant is that Eve measures again the
photon on the backward path after Alice's encoding, to capture the
information. Since Eve did know the state entering Alice's box she
can extract the information without perturbing the state. After
that she forwards the photon in the correct state to Bob. We call
this first PNS attack PNS$_{\mathcal{M}}$.

A second attack is more peculiar to LM05. Suppose that $n=2$ and
call the two photons in the pulse $p_{1}$ and $p_{2}$. As before
Eve can know the number of photons per pulse through a QND
measure. When $n<2$ Eve blocks the pulses. When $n=2$ she stores
$p_{1}$ and forwards $p_{2}$ to Alice; this let her remain
undetected during a possible CM on the forward path. On the way
back Eve captures again $p_{2}$. To gain Alice's information she
must decide whether the polarizations of $p_{1}$ and $p_{2}$ are
parallel or antiparallel: in the first case she would deduce the
logical value `$0$'; in the second case she would deduce `$1$'.
However, the discrimination between parallel and antiparallel
spins is not as simple as it appears at a first glimpse: while the
parallel-spin-state $\vert P\rangle =\vert \psi\rangle
_{p_{1}}\vert \psi\rangle _{p_{2}}$ is symmetric, the
antiparallel-spin-state $\vert AP\rangle =\vert \psi \rangle
_{p_{1}}\vert \psi^{\perp}\rangle _{p_{2}}$ is neither symmetric
nor antisymmetric. Upon symmetrizing $\vert AP\rangle $ we can
realize that it is not orthogonal to $\vert P\rangle $, and by
consequence it is not perfectly distinguishable from it (we remand
to~\cite{Bar04}, \cite{Pry05} for a complete treatment of this
problem). Actually an optimal measurement $\mathcal{M^\prime}$ is
a nonlocal one and gives Eve a conclusive result (between $\vert
P\rangle $ and $\vert AP\rangle $) with a probability $1/4$
\cite{Bar04}. Hence Eve can block all the ``inconclusive'' pulses
to gain full information and still remain undetected. However it
remains open the question of which photon must Eve forward to Bob:
upon obtaining this result, Eve does not know whether to give Bob
the state $\vert \psi\rangle $ or the state $\vert
\psi^{\perp}\rangle $, because she ignores the absolute value of
$\psi $\ prepared by Bob. This shows that two photons are not
sufficient for a perfect eavesdropping with $\mathcal{M}^\prime$.
Yet the complete attack can be accomplished with an additional
photon $p_{3}$: Eve should store $p_{3}$, execute
$\mathcal{M^\prime}$, and eventually encode $p_{3}$ according to
the conclusive outcome of $\mathcal{M^\prime}$; the photon
prepared in this way can be forwarded to Bob without risk of
detection.

The above analysis establishes that a perfect eavesdropping can be
realized with at least three photons in a pulse. It also
establishes that the measurement $\mathcal{M}$ represents a more
powerful resource for Eve than $\mathcal{M^\prime}$, for a number
of reasons: it gives information on the complete state $\psi$ of
the photons, not only on Alice's operation; the probability of
conclusive results is $1/2$ rather than $1/4$; Eve knows about the
conclusiveness of her measurement immediately, rather than after
Alice's encoding, and can use this information to improve her
strategy. For these reasons we only consider the robustness of
LM05 against the PNS$_{\mathcal{M}}$ attack
following~\cite{Luc06}.

{\gen In case of PNS attacks it was shown in~\cite{Brassard2000}
that a communication is secure until the signal sent by Alice
arrives at Bob's with a probability higher than the probability to
generate a multi-photon. In our present case the signal is the raw
gain $G_{raw}$, previously introduced for BB84,
Eq.\eqref{eq:GrawBB84}, and LM05, Eq.\eqref{eq:GrawLM05}, where we
have neglected the contribution of the dark counts.}
Hence, for BB84, the transmission is secure when the following
condition is fulfilled:
\begin{equation}\label{eq:SecRegBB84}
D^{BB84}=G_{raw}^{BB84}-P_{PNS}^{BB84}>0,
\end{equation}
with $G_{raw}^{BB84}$ defined by Eq.\eqref{eq:GrawBB84} and
\[
P_{PNS}^{BB84}=\sum_{i=2}^{\infty}e^{-\mu}\frac{\mu^{i}}{i!}=1-e^{-\mu}\left(
1+\mu\right)
\]
the probability that Alice accidentally emits 2 or more photons in
a single pulse.

All the same, for the LM05 we have:
\begin{equation}\label{eq:SecRegLM05}
D^{LM05}=G_{raw}^{LM05}-P_{PNS}^{LM05}>0,
\end{equation}
with $G_{raw}^{LM05}$ defined by Eq.\eqref{eq:GrawLM05} and
\[
P_{PNS}^{LM05}=\left(
\sum_{i=3}^{\infty}e^{-\mu}\frac{\mu^{i}}{i!}\right)
-\frac{1}{2}\frac{\mu^{3}}{6}e^{-\mu}=1-e^{-\mu}\left(  1+\mu+\frac{\mu^{2}%
}{2}+\frac{1}{2}\frac{\mu^{3}}{6}\right)
\]
the probability that Alice emits a number of photons dangerous for
LM05, as discussed above.

\begin{figure}[h!]
\includegraphics[width=12.0cm ]{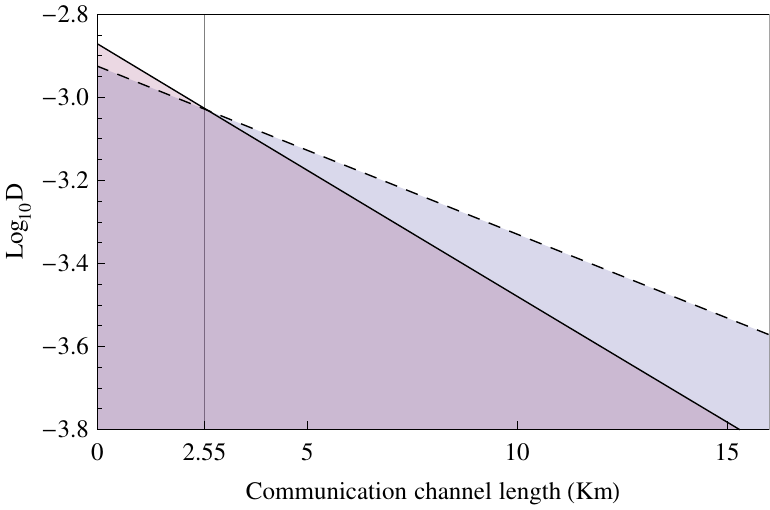}
\caption{Plot of the security regions of BB84 ($D^{BB84}$,
Eq.\eqref{eq:SecRegBB84}, from the dashed line to the axis of
abscissae) and LM05 ($D^{LM05}$, Eq.\eqref{eq:SecRegLM05}, from
the solid line to the axis of abscissae) versus the length of the
communication channel, in case of PNS attack by Eve. The BB84 is
superior to LM05 for longer distances, but LM05 is superior on
shorter distances. This is due to the higher resistance of LM05 to
PNS-like attacks.} \label{fig:PNS}
\end{figure}

Then we plot in Fig.~\ref{fig:PNS} $\log_{10} D^{BB84}$ and
$\log_{10} D^{LM05}$ and obtain a comparison of the security
regions pertaining to BB84 and LM05.
Differently from the BS attack, for the PNS attacks the BB84 is
more performant than LM05 on longer distances, higher than 2.55~Km
with our parameters, but LM05 is more performant on shorter
distances. Hence, while losses remain an important limiting factor
for any two-way quantum communication, the higher protection from
PNS attacks of the LM05 with respect to BB84 could enable an
advantage in terms of the secure rate for small-range setups like
the one in~\cite{DGH+06}.

\section{Conclusion}

In the present work we have described and characterized a
practical version of the two-way QKD protocol LM05 by considering
the most relevant {\tre single-particle} eavesdropping strategies
against it.

{\gen The security of the scheme is guaranteed by two QBERs, $q_1$
and $Q_{AB}$, which are not trivially related one to each other.
By assuming a noise model where $q_1 = Q_{AB}$ it turns out that
the most powerful attack against LM05 in \textit{direct
reconciliation} is the NORT, which provides an upper bound to
secure QKD equal to $q_1 = Q_{AB} = 10.0\%$, while the most
powerful attack in \textit{reverse reconciliation} is the
DCNOT$^*$, which provides a threshold of $q_1 = Q_{AB} = 11.9\%$.
The adopted noise model is based on the experimental evidence that
one-way and two-way channels can be controlled with the same level
of precision.
%
%
However other noise models can be considered as well, for instance
one in which $q_1 = 15\%$ and $Q_{AB} = 0$. In this case, if Eve
performs individual attacks only, it is not possible to distill a
secure key with the BB84 protocol, while it is possible with the
LM05. This in our opinion underlines the intrinsic difference
between the two ways of performing QKD and encourages further
research in this area. Moreover it is worth recalling that in the
present version of LM05 we did not include a QBER test on the
backward path of the communication channel; such an option, though
not very practical, can improve the obtained secrecy capacities.

In the limited scenario of single-particle attacks, we have
compared the two-way LM05 protocol with the standard BB84, which
uses a one-way quantum channel. We have found that the main
limitation of two-way QKD is the high loss-rate, even if it is
partially counter-balanced by the higher resistance to PNS-like
attacks. This entails that the field of application of two-way QKD
shall be limited to short communication channels, like those
typical of a QKD network, where the optimal average distance is
about 17~Km~\cite{ARD+09}.
Even so a conclusive answer about the two modes of performing QKD
can come only from a quantitative unconditional security proof for
two-way QKD, which is still lacking at present. In this respect,
it is our opinion that specific tools should be developed for this
task, e.g. an entanglement-based description specific to two-way
QKD. In fact, any argument too closely related to one-way QKD
fails to capture all the peculiarities of two-way QKD, like the
encoding on operators rather than on states
and the reverse reconciliation of the key.

While we are aware that further studies are {\tre needed} to make
the LM05 protocol and two-way QKD in general viable for
applications, we are also confident that the multi-way use of a
quantum channel could play an important role in future
developments of quantum cryptography.

\section*{Acknowledgments} Fruitful discussions with
N.~L\"{u}tkenhaus, G. Di~Giuseppe and R.~Kumar are acknowledged.
This work has been partly supported by the European Commission
under the Integrated Project ``Qubit APplications'' QAP funded by
the IST, Contract No. 015848 and partly by those Italian citizens
who assigned the $5\tcperthousand$~of their income tax to the
University of Camerino.



\begin{thebibliography}{999}
\bibitem{footnote3} To Eve it does not really matter the choice of the basis until
it is chosen on the equator of the Poincar\'{e} sphere. In fact
all the ``equatorial states'' are flipped by Alice's operation
i\^{Y}, so they provide her with full information. Moreover the
noise introduced on the forward path does not depend on the basis
choice, as shown in~\cite{Gisin02}.
\bibitem{NoteUncSec}
To the best of our knowledge an unconditional security proof for
two-way QKD is not yet available, although some authors used the
arguments of Ref.~\cite{CL04} to claim that.
\end{thebibliography}
\end{document}